
%
%

\def\m@th{\mathsurround=0pt}

\def\fsquare(#1,#2){
\hbox{\vrule$\hskip-0.4pt\vcenter to #1{\normalbaselines\m@th
\hrule\vfil\hbox to #1{\hfill$\scriptstyle #2$\hfill}\vfil\hrule}$\hskip-0.4pt
\vrule}}

\def\addsquare(#1,#2){\hbox{$
	\dimen1=#1 \advance\dimen1 by -0.8pt
	\vcenter to #1{\hrule height0.4pt depth0.0pt%
	\hbox to #1{%
	\vbox to \dimen1{\vss%
	\hbox to \dimen1{\hss$\scriptstyle~#2~$\hss}%
	\vss}%
	\vrule width0.4pt}%
	\hrule height0.4pt depth0.0pt}$}}

\def\Addsquare(#1,#2){\hbox{$
	\dimen1=#1 \advance\dimen1 by -0.8pt
	\vcenter to #1{\hrule height0.4pt depth0.0pt%
	\hbox to #1{%
	\vbox to \dimen1{\vss%
	\hbox to \dimen1{\hss$~#2~$\hss}%
	\vss}%
	\vrule width0.4pt}%
	\hrule height0.4pt depth0.0pt}$}}

\def\Fsquare(#1,#2){
\hbox{\vrule$\hskip-0.4pt\vcenter to #1{\normalbaselines\m@th
\hrule\vfil\hbox to #1{\hfill$#2$\hfill}\vfil\hrule}$\hskip-0.4pt
\vrule}}

\def\naga{%
	\hbox{$\vcenter to 0.4cm{\normalbaselines\m@th
	\hrule\vfil\hbox to 1.2cm{\hfill$\cdots$\hfill}\vfil\hrule}$}}

\def\Flect(#1,#2,#3){
\hbox{\vrule$\hskip-0.4pt\vcenter to #1{\normalbaselines\m@th
\hrule\vfil\hbox to #2{\hfill$#3$\hfill}\vfil\hrule}$\hskip-0.4pt
\vrule}}

\def\PFlect(#1,#2,#3){
\hbox{$\hskip-0.4pt\vcenter to #1{\normalbaselines\m@th
\vfil\hbox to #2{\hfill$#3$\hfill}\vfil}$\hskip-0.4pt}}

\dimen1=0.5cm\advance\dimen1 by -0.8pt

\def\vnaka{\normalbaselines\m@th\baselineskip0pt\offinterlineskip%
	\vrule\vbox to 0.6cm{\vskip0.5pt\hbox to \dimen1{$\hfil\vdots\hfil$}\vfil}\vrule}

\dimen2=1.5cm\advance\dimen2 by -0.8pt

\def\vnakal{\normalbaselines\m@th\baselineskip0pt\offinterlineskip%
	\vrule\vbox to 1.2cm{\vskip7pt\hbox to \dimen2{$\hfil\vdots\hfil$}\vfil}\vrule}

\hsize=13cm
\magnification=\magstep1
%
%

\vskip2.5cm
\centerline{\bf Quantum Jacobi-Trudi and Giambelli Formulae}
\centerline{\bf for $U_q(B^{(1)}_r)$ from Analytic Bethe Ansatz}
\vskip1.0cm \centerline{by}
\vskip1.0cm
\centerline{Atsuo Kuniba\footnote\dag{
E-mail: atsuo@hep1.c.u-tokyo.ac.jp}}
\centerline{Institute of Physics, University of Tokyo}
\centerline{Komaba 3-8-1, Meguro-ku, Tokyo 153 Japan}
\par\vskip0.3cm
\centerline{Yasuhiro Ohta\footnote\ddag{
E-mail: ohta@kurims.kyoto-u.ac.jp}}
\centerline{Faculty of Engineering, Hiroshima University}
\centerline{Higashi Hiroshima, Hiroshima 724 Japan}
\par\vskip0.3cm
\centerline{and}\par\vskip0.3cm
\centerline{Junji Suzuki\footnote\P{
E-mail: jsuzuki@tansei.cc.u-tokyo.ac.jp}} 
\centerline{Institute of Physics, University of Tokyo}
\centerline{Komaba 3-8-1, Meguro-ku, Tokyo 153 Japan}
\vskip5.0cm
\centerline{\bf Abstract}
\vskip0.2cm
\par
Analytic Bethe ansatz is executed for a wide class of
finite dimensional $U_q(B^{(1)}_r)$ modules.
They are labeled by skew-Young diagrams which, in general, 
contain a fragment corresponding to the spin representation.
For the transfer matrix spectra of the relevant vertex models,
we establish a number of formulae,
which are $U_q(B^{(1)}_r)$ analogues of the 
classical ones due to Jacobi-Trudi and Giambelli on
Schur functions.
They yield a full solution to the previously proposed 
functional relation ($T$-system), which is a Toda equation
on discrete space-time.

\vfill
\eject

\beginsection 1. Introduction

\noindent
In [KS1] analytic Bethe ansatz was worked out
for all the fundamental representations of the Yangians 
$Y(X_r)$ of classical types 
$X_r = B_r, C_r$ and $D_r$.
Namely, for any $a \in \{1,2, \ldots, r \}$,
a rational function $\Lambda^{(a)}_1(u)$ of the 
spectral parameter $u$ has been constructed, which should
describe the spectrum of the transfer matrices
of the corresponding solvable vertex models.
It is a Yangian analogue of the
character of the auxiliary space and satisfies a couple
of conditions required for it.
In particular $\Lambda^{(a)}_1(u)$ has been shown pole-free
provided that the Bethe ansatz equation (BAE) holds.
These results are also valid for $U_q(X^{(1)}_r)$ case
after replacing the rational functions by their natural 
$q$-analogues.
See [R,KS1] for general accounts on the 
analytic Bethe ansatz.
\par
In this paper we extend such analyses 
beyond the fundamental representations for $X_r = B_r$.
We introduce skew-Young diagrams $\lambda \subset \mu$ [M]
and a set of tableaux on them obeying a certain
semi-standard like conditions.
Then we construct the corresponding function
$T_{\lambda \subset \mu}(u)$ in terms of a sum over such 
tableaux via a certain rule.
The $T_{\lambda \subset \mu}(u)$ is to be regarded 
as the spectrum of the commuting
transfer matrix with
auxiliary space 
labeled by $\lambda \subset \mu$.
It has a dressed vacuum form (DVF) in the analytic Bethe ansatz. 
We shall rewrite $T_{\lambda \subset \mu}(u)$ 
in several determinantal forms,
where the matrix elements are only those $T_\mu(u)$
for the usual Young diagrams 
with shapes $\mu = (1^a), (m)$ or $(m+1,1^a)$.
They can be viewed as $U_q(B^{(1)}_r)$ analogues
of the classical Jacobi-Trudi and Giambelli formulae
on Schur functions [M].
Pole-freeness of the $T_{\lambda \subset \mu}(u)$ 
under BAE follows immediately from these formulae and our 
previous proof for the case $\mu = (1^a)$ [KS1].
These results correspond to
the case where the auxiliary space is even with 
respect to the tensor degree of the spin
representation.
We shall simply refer to such a case 
spin-even and spin-odd otherwise.
See the remark after (3.12) for a precise definition.
We will also treat the spin-odd case by 
using a modified skew-Young diagrams and 
semi-standard like conditions on them.
Combining these results, we obtain a full solution in terms of the DVF
to the transfer matrix functional relation ($T$-system)
proposed in [KNS].
This substantially achieves our program raised 
in [KS1] for $B_r$.
\par
A natural question here is, what is the finite dimensional 
auxiliary space labeled by those skew-Young diagrams as 
a representation space of $U_q(B^{(1)}_r)$ or $Y(B_r)$?
We suppose that it is an irreducible 
one in view that
all the terms in $T_{\lambda \subset \mu}(u)$ are
coupled to make the associated poles suprious 
under BAE.
Moreover we specify, in the Yangian 
context, the Drinfeld polynomial explicitly
based on some empirical procedure.
We shall also determine how the irreducible $Y(B_r)$ module
decomposes as a $B_r$ module through the embedding 
$B_r \hookrightarrow Y(B_r)$ for the spin-even case.
\par
The paper is organized as follows.
In the next section we recall the results 
in [KS1] on $U_q(B^{(1)}_r)$.
We then introduce the basic functions
$T^a(u)$ and $T_m(u)$ for all $a, m \in {\bf Z}_{\ge 0}$.
These are analogues of $a$-th anti-symmetric and 
$m$-th symmetric fusion transfer matrices (or its
eigenvalues), respectively.
For $1 \le a \le r-1$, we have 
$T^a(u) = \Lambda^{(a)}_1(u) = T_{(1^a)}(u)$ 
in the above.
The introduction of $T^a(u)$ with $a \ge r$ is a key
in this paper and we point out 
a new functional relation (2.14) among them.
In section 3 and 4, we treat the spin-even and odd cases,
respectively.
In terms of the DVFs in these sections, we give,
in section 5, a full solution
to the $T$-system [KNS] 
with an outline of the proof.
Until this point we will exclusively consider the situation
where the quantum space is formally trivial.
This means that the vacuum part in DVF is
always 1 as well as the ``left hand side'' of the BAE.
Section 6 includes a discussion on how to recover
the vacuum part for the non-trivial quantum spaces.
A prototype of them is a tensor product of 
irreducible finite dimensional modules such as (6.1).
The problem is essentially equivalent to 
specifying the left hand side of the BAE 
(cf section 2.4 in [KS1]) for such a general 
quantum space.
For the Yangian $Y(X_r)$, 
we propose quite generally for any $X_r$ that it is just given by 
a ratio of the relevant Drinfeld polynomials.\footnote\dag{
We thank E.K. Sklyanin and V.O. Tarasov for a discussion on this point.}
See (6.2).
Then we shall briefly indicate a way to recover the vacuum parts.
\par
Many formulae in section 3 are formally valid also 
for $U_q(A^{(1)}_r)$ under a suitable condition.
In particular $\lambda = \phi$ case of (3.5) 
has appeared in [BR], for which a representation theoretical
background is available in [C].
\par
We hope to report similar results for $C_r$ and $D_r$ 
cases in near future.

\beginsection 2. Review of the results on fundamental representations

\noindent
Here we shall recall the $B_r$ case of the results in [KS1].
Let $\{ \alpha_1, \ldots, \alpha_r \}$
and $\{ \Lambda_1, \ldots, \Lambda_r \}$ be the set of 
the simple roots and 
fundamental weights of $B_r$ ($r \ge 2$).
Our normalization is 
$t_1 = \cdots = t_{r-1} = {1 \over 2} t_r = 1$ for 
$t_a = 2/(\alpha_a \vert \alpha_a)$.
Then $(\alpha_a \vert \alpha_b) = 
{2 \over t_a}\delta_{a, b} - \delta_{a, b-1} - \delta_{a, b+1}$
and $(\alpha_a \vert \Lambda_b) = \delta_{a b}/{t_a}$.
The $U_q(B^{(1)}_r)$ BAE for the trivial quantum space 
reads [RW]
$$\eqalignno{
-1 &= \prod_{b = 1}^r 
{Q_b(v^{(a)}_k + (\alpha_a \vert \alpha_b)) \over
 Q_b(v^{(a)}_k - (\alpha_a \vert \alpha_b)) }\quad
\hbox{ for } 1 \le a \le r, \, 1 \le k \le N_a,
&(2.1)\cr
Q_a(u) &= \prod_{j=1}^{N_a} [u - v^{(a)}_j], 
&(2.2)\cr}
$$
where 
$[u] = (q^u - q^{-u})/(q-q^{-1})$
and $N_1, \ldots, N_r$ are some positive integers.
Throughout the paper we assume that $q$ is generic.
The LHS of (2.1) is just -1 as opposed to the non-trivial quantum space
case (6.2), which will be discussed in section 6.
Until then we shall focus on the dress parts 
in the analytic Bethe ansatz.
\par
Following [KS1] we introduce the set $J$ and the order 
$\prec$ in it as
$$\eqalignno{
&J = \{1, 2, \ldots, r, 0, {\bar r}, \ldots, {\bar 1} \},
&(2.3{\rm a})\cr
&1 \prec 2 \prec \cdots, \prec r \prec 0 \prec 
{\bar r} \prec \cdots, \prec {\bar 1}.
&(2.3{\rm b})\cr}
$$
For $a \in J$, define the function
$z(a; u)$ by
$$\eqalign{
z(a; u)  &=
      {{Q_{a-1}( u+a+1  ) 
            Q_{a}(u+a-2 )}\over
       { Q_{a-1}(u+a-1)Q_{a}(u+a)}} 
  \qquad 1\le a \le r,\cr
z(0; u)  &= 
    {{Q_r(u+r-2) Q_{r}(u+r+1)}\over
   { Q_{r}(u+r)Q_{r}(u+r-1)}},  \cr
z({\bar a}; u)  &= 
      {{Q_{a-1}( u+2r-a-2  ) 
            Q_{a}(u+2r-a+1 )}\over
       { Q_{a-1}(u+2r-a)Q_{a}(u+2r-a-1)}} 
  \qquad 1\le a \le r,\cr
}\eqno(2.4)$$
where we have set $Q_0(u) = 1$.
$z(a, u)$ is the dress part of the box 
$\Fsquare(0.5cm,a)$ in (4.4a) of [KS1], which corresponds
to a weight in the vector representation.
For $(\xi_1, \ldots, \xi_r) \in \{\pm \}^r$,
define the function $sp(\xi_1, \ldots, \xi_r; u)$
by the following recursion relation with
respect to $r$ and the initial condition $r = 2$.
$$\eqalign{
sp(+,+,\xi_3,\ldots,\xi_r; u) &= \tau^Q
sp(+,\xi_3,\ldots,\xi_r; u),\cr
sp(+,-,\xi_3,\ldots,\xi_r; u) &= 
{Q_1(u+r-{5\over 2})\over Q_1(u+r-{1\over 2})}
\tau^Q sp(-,\xi_3,\ldots,\xi_r; u),\cr
sp(-,+,\xi_3,\ldots,\xi_r; u) &= 
{Q_1(u+r+{3\over 2})\over Q_1(u+r-{1\over 2})}
\tau^Q sp(+,\xi_3,\ldots,\xi_r; u+2),\cr
sp(-,-,\xi_3,\ldots,\xi_r; u) &= \tau^Q
sp(-,\xi_3,\ldots,\xi_r; u+2).\cr}
\eqno(2.5{\rm a})$$
$$\eqalign{
sp(+,+; u)
   &={Q_2(u-{1\over 2}) \over Q_2(u+{1\over 2})},\cr
sp(+,-; u)
   &={Q_1(u-{1\over 2})Q_2(u+{3\over 2}) \over 
      Q_1(u+{3\over 2})Q_2(u+{1\over 2})},\cr
sp(-,+; u)
   &={Q_1(u+{7\over 2})Q_2(u+{3\over 2}) \over 
      Q_1(u+{3\over 2})Q_2(u+{5\over 2})},\cr
sp(-,-; u)
   &={Q_2(u+{7\over 2}) \over Q_2(u+{5\over 2})}.\cr}
\eqno(2.5{\rm b})
$$
In (2.5a) $\tau^Q$ is the operation $Q_a \rightarrow Q_{a+1}$, namely,
$$\eqalign{
\tau^Q & F(Q_1(u+x^1_1), Q_1(u+x^1_2), \ldots,
         Q_2(u+x^2_1), Q_2(u+x^2_2), \ldots)  \cr
&=  F(Q_2(u+x^1_1), Q_2(u+x^1_2), \ldots,
         Q_3(u+x^2_1), Q_3(u+x^2_2), \ldots)}
\eqno(2.6)
$$
for any function $F$.
$sp(\xi_1, \ldots, \xi_r; u)$ is the dress part of the box 
$\overbrace{
\Flect(0.4cm,2.2cm,{\xi_1,\xi_2, \cdots , \xi_r})}^{r}$
in (4.25,26) of [KS1].
\par
Now we introduce the meromorphic 
functions $T^a(u)$ and $T_m(u)$ of $u$ 
for any $a, m \in {\bf Z}_{\ge 0}$
by the following ``non-commutative generating series''
$$\eqalignno{
&(1+z({\bar 1};u)X) \cdots (1+z({\bar r};u)X)
(1-z(0;u)X)^{-1}(1+z(r;u)X) \cdots (1+z(1;u)X)\cr
& = \sum_{a=0}^\infty T^a(u+a-1) X^a, &(2.7{\rm a})\cr
&(1-z(1;u)X)^{-1} \cdots (1-z(r;u)X)^{-1}
(1+z(0;u)X)
(1-z({\bar r};u)X)^{-1} \cdots (1-z({\bar 1};u)X)^{-1}\cr
& = \sum_{m=0}^\infty T_m(u+m-1) X^m, &(2.7{\rm b})\cr}
$$
where $X$ is a difference operator with the commutation relation
$$
X Q_a(u) = Q_a(u+2) X \quad \hbox{ for any } 1 \le a \le r.
\eqno(2.8)
$$
Thus $X z(a;u) = z(a; u+2) X$ for any $a \in J$.
We set $T^a(u) = T_m(u) = 0$ for $a, m < 0$.
An immediate consequence of the above definition is
$$\eqalignno{
\delta_{i j} &= \sum_{k=0}^N (-)^{i-k}T_{i-k}(u+i+k)
T^{k-j}(u+k+j) &(2.9{\rm a})\cr
&= \sum_{k=0}^N (-)^{i-k}T_{i-k}(u-i-k)
T^{k-j}(u-k-j) &(2.9{\rm b})\cr}
$$
for any $N \ge 0$ and $0 \le i, j \le N$.
Define $T^{(a)}_1(u)$ for $1 \le a \le r$ by
$$\eqalign{
T^{(a)}_1(u) &= T^a(u) \quad \hbox{ for } 1 \le a \le r-1,\cr
T^{(r)}_1(u) &= \sum_{\xi_1, \ldots, \xi_r = \pm }
sp(\xi_1, \ldots, \xi_r; u).\cr}
\eqno(2.10)
$$
Then $T^{(a)}_1(u)$ coincides with the dress part of $\Lambda^{(a)}_1(u)$
in [KS1] for all $1 \le a \le r$.
\proclaim Theorem 2.1. 
$T^{(r)}_1(u), T^a(u)$ and
$T_m(u) (\forall a, m \in {\bf Z})$ are pole-free
provided that the BAE (2.1) holds.
\par\noindent
For $T^{(r)}_1(u)$ and 
$T^a(u)$ with $a \le r-1$, this was proved in [KS1] in 
the more general setting
including the vacuum parts.
The other cases can be verified quite similarly.
$T^{(1)}_1(u)$ and $T^{(r)}_1(u)$ was considered earlier [R].
\par
The functions $z(a; u)$ and 
$sp(\xi_1, \ldots, \xi_r; u)$ are related as follows.
Given two sequences
$(\xi_1, \ldots, \xi_r)$ and 
$(\eta_1, \ldots, \eta_r) \in \{ \pm \}^r$, 
we define $i_1 < \cdots < i_k, I_1 < \cdots < I_{r-k}\, (0 \le k \le r)$ 
and 
$j_1 < \ldots < j_l, J_1 < \ldots < J_{r-l}\, (0 \le l \le r)$ 
by the following.
$$\eqalign{
&\xi_{i_1} = \cdots = \xi_{i_k} = +, \,
 \xi_{I_1} = \cdots = \xi_{I_{r-k}} = -,\cr
&\eta_{j_1} = \cdots = \eta_{j_l} = -, \,
 \eta_{J_1} = \cdots = \eta_{J_{r-l}} = +.\cr}
\eqno(2.11)
$$
Then we have
\proclaim Proposition 2.2.
For any $a \in {\bf Z}_{\ge 0}$, 
$$\eqalign{
&sp(\xi_1, \ldots, \xi_r; u-r+a+{1 \over 2})
sp(\eta_1, \ldots, \eta_r; u+r-a-{1 \over 2})\cr
&= \prod_{n=1}^a z(b_n; u+a+1-2n)\quad\quad  \hbox{ if } k + l \le a,\cr}
\eqno(2.12{\rm a})
$$
where
$$b_n = \cases{i_n & for $1 \le n \le k$\cr
               0 &   for $k < n \le a-l$\cr
               \overline{j_{a+1-n}} & for $a-l < n \le a$ \cr}.
\eqno(2.12{\rm b})
$$
For any $a \in {\bf Z}_{\le 2r-1}$, 
$$\eqalign{
&sp(\xi_1, \ldots, \xi_r; u-r+a+{1\over 2})
sp(\eta_1, \ldots, \eta_r; u+r-a-{1\over 2})\cr
&= \prod_{n=1}^{2r-1-a}
z(b^\prime_n; u+2r-a-2n)\quad \quad\hbox{ if } k + l \ge a+1,\cr}
\eqno(2.13{\rm a})
$$
where
$$b^\prime_n = \cases{J_n & for $1 \le n  \le r-l$\cr
               0 &   for $r-l < n \le r+k-1-a$\cr
               \overline{I_{2r-a-n}} & for $r+k-1-a < n \le 2r-1-a$ \cr}.
\eqno(2.13{\rm b})
$$
\par
This enables the evaluation of the product
$sp(\xi_1, \ldots, \xi_r; u-r+a+{1\over 2})
sp(\eta_1, \ldots, \eta_r; u+r-a-{1\over 2})$
for any $\{ \xi_i \}, \{ \eta_i \}$ and $a \in {\bf Z}$
in terms of $z$ (2.4).
For $1 \le a \le r-1$, (2.12) is theorem A.1 in [KS1].
It is straightforward to extend it to any $a \in {\bf Z}_{\ge 0}$.
Eq. (2.13) can be derived from (2.12) by replacing
$a$ by $2r-1-a$.
Note in (2.12b) that
$b_1 \prec \cdots \prec b_k \prec b_{k+1} = \cdots = 
b_{a-l} = 0 \prec b_{a-l+1} \prec \cdots \prec b_a \in J$.
A similar inequality holds also for $b^\prime_n$.
Comparing them with (2.7a) and (2.10) we get
\par\noindent
\proclaim Theorem 2.3.
$$
T^a(u) + T^{2r-1-a}(u) = T^{(r)}_1(u-r+a+{1 \over 2})
 T^{(r)}_1(u+r-a-{1 \over 2})\quad 
\forall a \in {\bf Z}.\eqno(2.14)
$$
\par
This is invariant under the exchange
$a \leftrightarrow 2r-1-a$.
If $a < 0$ or $a > 2r-1$, there is in fact only one term
on the LHS.
The new functional relation (2.14) will play an important role
in this paper.
It is also valid
after including the vacuum parts.
See section 6.

\beginsection 3. Spin-even case

\noindent
Let $\mu = (\mu_1, \mu_2, \ldots)$, $\mu_1 \ge \mu_2 \ge \cdots \ge 0$
be a Young diagram and 
$\mu^\prime = (\mu^\prime_1, \mu^\prime_2, \ldots)$ be its transpose.
We let $d_\mu$ denote the length of the main diagonal of $\mu$.
By a skew-Young diagram we mean a pair of 
Young diagrams $\lambda \subset \mu$.
It is depicted by the region corresponding to the 
subtraction $\mu -\lambda$.
See the Fig.3.1 for example.
\par
\centerline{Fig.3.1}
\par
For definiteness, we assume that 
$\lambda^\prime_{\mu_1} = \lambda_{\mu^\prime_1} = 0$.
A Young diagram $\mu$ is naturally identified with a skew-Young diagram 
$\phi \subset \mu$.
By an {\it admissible} tableau $b$ 
on a skew-Young diagram $\lambda \subset \mu$ we mean an assignment
of an element $b(i,j) \in J$ to 
the $(i,j)$-th box in $\lambda \subset \mu$
under the following rule:
(We locate $(1,1)$ at the top left corner of 
$\mu$, $(i+1,j)$ and 
$(i,j+1)$ to the below and the right of $(i,j)$, 
respectively.)
$$\eqalign{
&b(i,j) \preceq b(i,j+1), \qquad b(i,j) \prec b(i+1,j)\quad
\hbox{with the exception that}\cr
&b(i,j) = b(i,j+1) = 0 \, \hbox{ is forbidden}, \quad
b(i,j) = b(i+1,j) = 0 \, \hbox{ is allowed}.\cr
}\eqno(3.1)
$$
Without the exception this coincides with
the usual definition of the semi-standard Young tableau.
Denote by $Atab(\lambda \subset \mu)$ the set of 
admissible tableaux on $\lambda \subset \mu$.
\par
Given a skew-Young diagram $\lambda \subset \mu$,
we define the function 
$T_{\lambda \subset \mu}(u)$ as the following 
sum over the admissible tableaux.
$$
T_{\lambda \subset \mu}(u) = \sum_{b \in Atab(\lambda \subset \mu)}
\prod_{(i,j) \in (\lambda \subset \mu)}
z(b(i,j); u+\mu^\prime_1 - \mu_1 - 2i+2j).
\eqno(3.2)
$$
Comparing this with (2.7) we have
$$\eqalignno{
T^a(u) &= T_{(1^a)}(u): \hbox{ single column of length } a,&(3.3{\rm a})\cr
T_m(u) &= T_{(m)}(u): \hbox{ single row of length } m.&(3.3{\rm b})\cr
}$$
We also prepare a notation for the single hook,
$$T_{k,l} = T_{(l+1,1^k)}(u). \eqno(3.3{\rm c})
$$
Our main result in this section is
\par
\proclaim Theorem 3.1.
$$
T_{\lambda \subset \mu}(u) = 
det \pmatrix{
0& \cdots & 0 & R_{1 1}& \cdots & R_{1 d_\mu}\cr
\vdots & \ddots & \vdots & \vdots &   & \vdots \cr
0& \cdots & 0 & R_{d_\lambda 1}& \cdots & R_{d_\lambda d_\mu}\cr
C_{1 1} & \cdots & C_{1 d_\lambda} & 
H_{1 1} & \cdots & H_{1 d_\mu} \cr
\vdots &  & \vdots & \vdots & \ddots & \vdots \cr
C_{d_\mu 1} & \cdots & C_{d_\mu d_\lambda} & 
H_{d_\mu 1} & \cdots & H_{d_\mu d_\mu} \cr},\eqno(3.4{\rm a})
$$
where 
$$\eqalign{
R_{i j} &= T_{\mu_j - \lambda_i +i-j}
(u+\mu^\prime_1-\mu_1 + \mu_j + \lambda_i -i-j+1),\cr
C_{i j} &= - T^{\mu^\prime_i - \lambda^\prime_j -i+j}
(u+\mu^\prime_1-\mu_1 - \mu^\prime_i - \lambda^\prime_j +i+j-1),\cr
H_{i j} &= T_{\mu^\prime_i -i, \mu_j-j}
(u+\mu^\prime_1-\mu_1 - \mu^\prime_i +\mu_j +i-j).\cr}
\eqno(3.4{\rm b})
$$
\par
Two particular cases corresponding to the formal choices
$\mu_i = \lambda_i$ or 
$\mu^\prime_i = \lambda^\prime_i$
for $1 \le i \le d_\lambda = d_\mu$ yield simpler formulae.
In these cases, redefining $\mu_i, \mu^\prime_i, 
\lambda_i$ and $\lambda^\prime_i$ so that
$\lambda^\prime_{\mu_1} = \lambda_{\mu^\prime_1}=0$,
we have
$$\eqalignno{
T_{\lambda \subset \mu}(u)
&= det_{1 \le i,j \le \mu_1}
(T^{\mu^\prime_i - \lambda^\prime_j -i+j}
(u+\mu^\prime_1-\mu_1 - \mu^\prime_i - \lambda^\prime_j +i+j-1)),
&(3.5{\rm a})\cr
&= det_{1 \le i,j \le \mu^\prime_1}
(T_{\mu_j - \lambda_i +i-j}
(u+\mu^\prime_1-\mu_1 + \mu_j + \lambda_i -i-j+1)).
&(3.5{\rm b})\cr}
$$
Eq.(3.5a) can be verified, for example, by induction on $\mu_1$, i.e.,
by showing the same recursive relation for the tableau sum (3.2)
as an expansion of the determinant.
Then (3.5b) follows from (2.9).
Theorem 3.1 is proved from these results by applying 
Sylvester's theorem on determinants.
From (3.5a) and Theorem 2.1 one has
\proclaim Corollary.
$T_{\lambda \subset \mu}(u)$ is pole-free provided the BAE (2.1) holds.
\par
The admissibility condition (3.1) leads to the above conclusion 
although it is by no means obvious in the defining expression (3.2).
Despite the exception in (3.1),
our formulae (3.4) and (3.5) formally coincide with 
the classical ones due to Giambelli and Jacobi-Trudi
on Schur functions [M] if one drops
the $u$-dependence (or in the limit 
$\vert u \vert \rightarrow \infty$).
If $\mu^\prime_{i+1} - \lambda^\prime_i > 2r$ for some $i$,
$Atab(\lambda \subset \mu) = \phi$.
Correspondingly, one can show that the determinant (3.5a) is
vanishing using the fact that $T^a(u)$ factorizes
for $a \ge 2r$ due to Theorem 2.3.
Henceforth we assume that 
$\mu^\prime_{i+1} - \lambda^\prime_i \le 2r$
for $1 \le i \le \mu_1$.
(We set $\mu^\prime_{\mu_1 + 1} = -\infty$.)
\par
The $T_{\lambda \subset \mu}(u)$ (3.2)
describes the spectrum of the transfer matrix 
whose auxiliary space is labeled by the skew-Young diagram
$\lambda \subset \mu$ and $u$.
Denote the space by $W_{\lambda \subset \mu}(u)$.
We suppose it is an irreducible finite dimensional module 
over $Y(B_r)$ (or $U_q(B^{(1)}_r)$ in the trigonometric case)
in view that all the terms in (3.2) 
seem coupling to make the apparent poles suprious under BAE. 
Now we shall specify the Drinfeld polynomial
$P_a(\zeta)$ [D] that characterizes 
$W_{\lambda \subset \mu}(u)$ based on some 
empirical procedure.
Our convention slightly differs
from the original one in Theorem 2 of [D]
in such a way that
$$
1 + \sum_{k=0}^\infty d_{i k} \zeta^{-k-1} = 
{P_i(\zeta + {1 \over t_i}) \over 
P_i(\zeta - {1 \over t_i})}.
\eqno(3.6)
$$
For any $b \in Atab(\lambda \subset \mu)$, the 
corresponding summand (3.2) has the form
$$
\prod_{a=1}^r
{Q_a(u+x^a_1) \cdots Q_a(u+x^a_{i_a}) \over
Q_a(u+y^a_1) \cdots Q_a(u+y^a_{i_a})}, \eqno(3.7)
$$
where $x^a_j, y^a_j$ and $i_a$ are specified from $b$.
This summand carries the $B_r$-weight
$$
wt(b) = \sum_{a=1}^r \Bigl(
{t_a \over 2} \sum_{j=1}^{i_a} (y^a_j - x^a_j) \Bigr) \Lambda_a
\eqno(3.8)
$$
in the sense that
$\lim_{q^u \rightarrow \infty}
\hbox{(3.7)} = q^{-2(wt(b) \vert \sum_{a=1}^r N_a \alpha_a)}$.
From $Atab(\lambda \subset \mu)$, take 
such $b_0$ that $wt(b_0)$ is highest,
which corresponds to the ``top term'' in section 2.4 of [KS1].
In our case, such $b_0$ is unique and given as follows.
Fill the left most column of $\lambda \subset \mu$
from the top to the bottom by assigning the first 
$\mu^\prime_1 - \lambda^\prime_1$ letters
from the sequence
$1,2,\ldots, r,0,0,\ldots$.
Given the $(i-1)$-th column, the $i$-th column is built 
from the top to the bottom
by taking the first 
$\mu^\prime_i - \lambda^\prime_i$ letters
from the sequence
$1,2,\ldots,r,\overbrace{0,\ldots,0}^k,\overline{r},
\overline{r-1},\ldots,\overline{1}$, where
$k = \hbox{max}(0,\hbox{min}(
\lambda^\prime_{i-1} - \lambda^\prime_i,
\mu^\prime_i - \lambda^\prime_i - r))$.
(We set $\lambda^\prime_0 = +\infty$.)
See the example in Fig.3.2.
\par
\centerline{Fig.3.2.}
\par\noindent
It turns out that (3.7) for the top term $b_0$ 
can be expressed uniquely in the form
$$
\prod_{a=1}^r \prod_{j=1}^{M_a}
{Q_a(u+z^a_j - {1 \over t_a}) \over
Q_a(u+z^a_j + {1 \over t_a})}\eqno(3.9)
$$
for some $M_a$ and $\{ z^a_j \vert 1 \le j \le M_a \}$
up to the permutations of $z^a_j$'s for each $a$.
We then propose that the Drinfeld polynomial 
$P^{W_{\lambda \subset \mu}(u)}_a(\zeta)$ for 
$W_{\lambda \subset \mu}(u)$ is given by
$$
P^{W_{\lambda \subset \mu}(u)}_a(\zeta) = \prod_{j=1}^{M_a}
(\zeta - u - z^a_j)\quad 1 \le a \le r.
\eqno(3.10)
$$
In our case, it reads explicitly as follows.
$$\eqalign{
P^{W_{\lambda \subset \mu}(u)}_a(\zeta)
&= \prod_{ \scriptstyle 1 \le i \le \mu_1  \atop
                \scriptstyle  \mu'_i-\lambda'_i=a }
            (\zeta-u-\mu'_1 +\mu_1+1+a+2\lambda'_i-2i)\cr
& \times \prod_{ \scriptstyle 1 \le i \le \mu_1-1 \atop
                \scriptstyle  \mu^\prime_{i+1}-\lambda'_i=2r-a }
            (\zeta-u-\mu'_1 +\mu_1+2+a+2\lambda'_i-2i)\quad 1 \le a \le r-1,\cr}
\eqno(3.11{\rm a})
$$
$$\eqalign{
P^{W_{\lambda \subset \mu}(u)}_r(\zeta)
&= \prod_{ \scriptstyle 1 \le i \le \mu_1  \atop
                \scriptstyle \lambda'_i+r \le \mu'_i \le \lambda'_{i-1}+r }
            (\zeta-u-\mu'_1 +\mu_1+2\mu'_i-2i-r+{3\over 2})\cr
& \times \prod_{ \scriptstyle 1 \le i \le \mu_1 \atop
                \scriptstyle  \mu^\prime_{i+1}\le \lambda'_i+r \le \mu'_i }
            (\zeta-u-\mu'_1 +\mu_1+2\lambda'_i-2i+r+{1\over 2}),\cr}
\eqno(3.11{\rm b})
$$
where we have set 
$$\mu'_{\mu_1+1}=-\infty, \quad \lambda'_0=\infty.
\eqno(3.12)$$
We will call the irreducible finite dimensional
$Y(B_r)$ module spin-even (resp. spin-odd)
if and only if the characterizing Drinfeld
polynomial $P_r(\zeta)$ is even (resp. odd) degree.
The one in (3.11b) is even for any skew-Young diagram
$\lambda \subset \mu$.
For example, in the case of the single 
column or row (3.3), (3.11) reads
$$\eqalignno{
P^{W_{(1^c)}(u)}_a(\zeta) &= \cases{
(\zeta - u)^{\delta_{a c}} & $1 \le c < r$\cr
\bigl( (\zeta - u + c - r + {1 \over 2})
       (\zeta - u - c + r - {1 \over 2}) \bigr)^{\delta_{a r}}
& $c \ge r$ \cr}, &(3.13{\rm a})\cr
P^{W_{(m)}(u)}_a(\zeta) &= 
\bigl( (\zeta - u+m-1)(\zeta - u+m-3) \cdots (\zeta - u -m+1)
\bigr)^{\delta_{a 1}}.&(3.13{\rm b})\cr}
$$
\par
As a $B_r$ module, the $Y(B_r)$ module 
$W_{\lambda \subset \mu}(u)$ decomposes as 
$$
W_{\lambda \subset \mu}(u) \simeq 
\sum_{\eta} \bigl(
\sum_{\kappa, \nu} LR^\mu_{\lambda \nu}
LR^{\nu}_{(2\kappa)^\prime \eta} \bigr)
\pi_{O(2r+1)}(V_\eta),\eqno(3.14)
$$
which is $u$-independent.
Here $LR^\mu_{\lambda \nu}$ etc denote the 
Littlewood-Richardson coefficients
for the universal character ring $\Lambda$ 
of $GL$ type introduced in [KT].
The sums run over all the Young diagrams $\eta, \nu$ and 
$\kappa = (\kappa_1, \kappa_2, \ldots)$, 
where $(2\kappa)^\prime$
stands for the transpose of
$2\kappa = (2\kappa_1, 2\kappa_2, \ldots)$.
$\pi_{O(2r+1)}(V_\eta)$ is the image of the 
specialization homomorphism [KT].
It is equal to ($\pm 1$ or 0) ``times'' 
the irreducible $B_r$ module $V_{\eta^\ast}$
with the highest weight labeled by the Young diagram
$\eta^\ast$ with $(\eta^{\ast})^ \prime_1 \le r$.
They are determined according to the equality
$\pi_{O(2r+1)}(\chi(\eta)) = (\pm 1 \hbox{ or } 0)
\times \chi(\eta^\ast)$ at the character level [KT].
%


\beginsection 4. Spin-odd case

\noindent
Consider the following subset 
$Spin \subset Atab((1^r))$.
$$
\hbox{
  \m@th\baselineskip0pt\offinterlineskip
   \vbox{ 
      \hbox{$\fsquare(0.5cm,\hbox{$i_1$})$}\vskip-0.4pt
      \hbox{$\vnaka$}\vskip-0.4pt
	     \hbox{$\fsquare(0.5cm,\hbox{$i_r$})$}\vskip-0.4pt
        }
      } 
\raise 4ex \hbox{$\in Spin \Leftrightarrow
\cases{
i_1 \prec \cdots \prec i_r \in J,\cr
0 \hbox{ is not contained},\cr
\hbox{only one of } i \hbox{ and } \overline{ i } \hbox{ is contained
for any } 1 \le i \le r.\cr}$}
\eqno(4.1)
$$
There is a bijection $\iota: Spin \rightarrow 
\{ (\xi_1, \ldots, \xi_r) \mid \xi_j = \pm \}$ sending
(4.1) with 
$1 \preceq i_1 \prec \cdots \prec i_k \preceq r \prec 
\overline{r} \preceq i_{k+1} \prec \cdots \prec i_r \preceq 
\overline{1}$ to
such $(\xi_1, \ldots, \xi_r)$ that
$\xi_{i_1} = \cdots = \xi_{i_k} = +,\,
\xi_{\overline{i_{k+1}}} = \cdots = \xi_{\overline{i_r}} = -$,
where we interpret $\overline{k} = i$ if $k = \overline{i}$.
Thus the latter of (2.10) can also be written as
$T^{(r)}_1(u) = \sum_{b \in Spin} sp(\iota(b);u)$.
This type of $\iota$ has also been utilized in [KN].
\par
For a skew-Young diagram $\lambda \subset \mu$ with
$\mu^\prime_1 - \lambda^\prime_1 \ge r$,
hatch the bottom $r$ boxes in the leftmost column, 
which we call an L-hatched skew-Young diagram
$\lambda \subset \mu$.
See Fig.4.1.
\par \centerline{Fig.4.1.}\par
\noindent
Consider a tableau $b$ on it, namely,
a map $b: \hbox{ L-hatched } \lambda \subset \mu 
\rightarrow J$.
We call a tableau $b$ on an L-hatched $\lambda \subset \mu$
{\it L-admissible} if and only if all of the following
three conditions are valid.
($n = \mu^\prime_2 - (\mu^\prime_1 - r)$ and see Fig. 4.2 
for the definitions of $i_l$ and $j_l$.)
$$\eqalign{
\hbox{(i)}& \hbox{ hatched part } \in Spin,
\, \hbox{ and } (3.1) \hbox{ for non-hatched part},\cr
\hbox{(ii)}&\, j_0 \prec i_1,\cr
\hbox{(iii)}&\, i_1 \preceq j_1, \ldots, i_n \preceq j_n 
\hbox{ or there exists } k \in \{1, \ldots, n \} \hbox{ such that }\cr
&\, i_1 \preceq j_1, \ldots, i_{k-1} \preceq j_{k-1} \hbox{ and }
\overline{r} \preceq j_k \prec i_k \preceq \overline{1}.\cr}
\eqno(4.2)
$$
Here (ii) is void when $\mu^\prime_1 - \lambda^\prime_1 = r$ and so is
(iii) for $n=0$.
\par \centerline{Fig.4.2.}\par
\noindent
Denote by $Atab_L(\lambda \subset \mu)$ the set of 
L-admissible tableaux on the L-hatched  $\lambda \subset \mu$.
We note that 
$Atab(\lambda \subset \mu) \not\subseteq
Atab_L(\lambda \subset \mu)$ nor
$Atab(\lambda \subset \mu) \not\supseteq
Atab_L(\lambda \subset \mu)$.
Given an L-hatched skew-Young diagram 
$\lambda \subset \mu$, we define the function
$S_{\lambda \subset \mu}^L(u)$ by 
$$\eqalign{
S_{\lambda \subset \mu}^L(u) &= 
\sum_{b \in Atab_L(\lambda \subset \mu)}
sp(\iota(\hbox{hatched part});u)\cr
&\times
\prod_{(i,j) \in \hbox{ non hatched part of } (\lambda \subset \mu)}
z(b(i,j); u+2\mu^\prime_1-r-2i+2j-{3 \over 2}).}\eqno(4.3)
$$
\par
We have an L $\leftrightarrow$ R  (left vs. right) dual of these 
definitions as follows.
For a skew-Young diagram $\lambda \subset \mu$ with
$\mu^\prime_{\mu_1} \ge r$ 
(remember we assumed $\lambda^\prime_{\mu_1}=0$),
hatch the top $r$ boxes in the rightmost column, 
which we call an R-hatched skew-Young diagram
$\lambda \subset \mu$.
See Fig.4.3.
\par \centerline{Fig.4.3.}\par
\noindent
Consider a tableau 
$b: \hbox{ R-hatched } \lambda \subset \mu 
\rightarrow J$.
We call a tableau $b$ on an R-hatched $\lambda \subset \mu$
{\it R-admissible} if and only if all of the following
three conditions are valid.
($n = r  - \lambda^\prime_{\mu_1 - 1}$ and see Fig. 4.4 
for the definitions of $i_l$ and $j_l$.)
$$\eqalign{
\hbox{(i)}&\, \hbox{ hatched part } \in Spin,
\, \hbox{ and } (3.1) \hbox{ for non-hatched part},\cr
\hbox{(ii)}&\, i_1 \prec j_0,\cr
\hbox{(iii)}&\, j_1 \preceq i_1, \ldots, j_n \preceq i_n 
\hbox{ or there exists } k \in \{1, \ldots, n \} \hbox{ such that }\cr
&\, j_1 \preceq i_1, \ldots, j_{k-1} \preceq i_{k-1} \hbox{ and }
1 \preceq i_k \prec j_k \preceq r,\cr}
\eqno(4.4)
$$
where (ii) is void when $\mu^\prime_{\mu_1} = r$ and so is
(iii) for $n=0$.
\par \centerline{Fig.4.4.}\par
\noindent
Denoting by $Atab_R(\lambda \subset \mu)$ the set of 
R-admissible tableaux on the R-hatched  $\lambda \subset \mu$,
we define 
$$\eqalign{
S_{\lambda \subset \mu}^R(u) &= 
\sum_{b \in Atab_R(\lambda \subset \mu)}
sp(\iota(\hbox{hatched part});u)\cr
&\times
\prod_{(i,j) \in \hbox{ non hatched part of } (\lambda \subset \mu)}
z(b(i,j); u-2\mu_1+r-2i+2j+{3 \over 2}).}\eqno(4.5)
$$
\par
Our first main results in this section is
\proclaim Theorem 4.1.
$$\eqalignno{
S_{\lambda \subset \mu}^L(u) 
&= det_{1 \le i,j \le \mu_1}({\cal S}^L_{i j}) &(4.6{\rm a})\cr
&= det_{1 \le i,j \le \mu^\prime_2}
(\overline{\cal S}^L_{i j}), &(4.6{\rm b})\cr}
$$
where
$$\eqalignno{
{\cal S}^L_{i j} &= 
  \cases{  T^{\mu^\prime_j-\lambda^\prime_i+i-j} 
(u+2\mu^\prime_1-\mu^\prime_j-\lambda^\prime_i+i+j-r-{5\over 2})
                \, \,  \, j \ge 2  \cr
           T^{(r)}_1(u+2i-2+2(\mu^\prime_1-\lambda^\prime_i-r)) 
                  \,\, \,  j=1 \cr
         },&(4.7{\rm a})\cr
\overline{\cal  S}^L_{i j} &= 
  \cases{  T_{\mu_i-\lambda_j-i+j} 
(u+2\mu^\prime_1+\mu_i+\lambda_j-i-j-r-{1\over2})
                   \,\,  1\le j \le \lambda'_1  \cr
           {\cal H}^L_{\mu_i+\lambda^\prime_1-i}
(u+2\mu^\prime_1-2\lambda'_1-2r) 
                   \, \,\, j=\lambda'_1+1      \cr
           T_{\mu_i-i+j-1} (u+2\mu^\prime_1+\mu_i-i-j-r+{1\over2})
                   \, \,\,  j> \lambda'_1+1 \cr
         },&(4.7{\rm b})\cr
{\cal H}^L_m(u) &= \sum_{l=0}^{m} (-1)^l
             T^{(r)}_1 (u+2l) T_{m-l}(u+m+r+l-{1 \over 2}).
&(4.7{\rm c})\cr}
$$
\par
From (4.6a), (4.7a,c) and Theorem 3.1,
${\cal H}^L_m(u)$ is equal to the L-hatched 
hook $S^L_{(m+1,1^{r-1})}(u)$.
\par
For an R-hatched diagram $\lambda \subset \mu$,
let $\xi \subset \eta$ be the sub-diagram obtained
by removing the rightmost column of $\lambda \subset \mu$.
See Fig. 4.5.
\par\centerline{ Fig. 4.5.} \par\noindent
Thus for example 
$\eta_i = \mu_1 -1$ for 
$1 \le i \le \mu^\prime_{\mu_1} - \lambda^\prime_{\mu_1 - 1}$.
Then another main result in this section is the R-hatched 
version of the previous theorem as follows.
\proclaim Theorem 4.2.
$$\eqalignno{
S_{\lambda \subset \mu}^R(u) 
&= det_{1 \le i,j \le \mu_1}({\cal S}^R_{i j}) &(4.8{\rm a})\cr
&= det_{1 \le i,j \le \eta^\prime_1}
(\overline{\cal S}^R_{i j}), &(4.8{\rm b})\cr}
$$
where
$$\eqalignno{
{\cal  S}^R_{i j} &= 
  \cases{T^{\mu^\prime_j-\lambda^\prime_i+i-j}
(u-2\mu_1-\mu^\prime_i-\lambda^\prime_j+i+j+r+{1\over 2})
                \,\,\,   j \le \mu_1-1  \cr
           T^{(r)}_1 (u-2\mu_1-2\mu^\prime_i+2i+2r) 
                  \,\,\,  j=\mu_1 \cr
         },&(4.9{\rm a})\cr
\overline{\cal S}^R_{i j} &= 
  \cases{  
T_{\eta_i-\xi_j-i+j}(u-2\lambda'_{\mu_1-1}-
                                2\eta_1+\eta_i+\xi_j-i-j+r+{1\over2})  
            \,       i\ne \eta'_1-\mu^\prime_1+\mu'_{\mu_1}    \cr
             {\cal H}^R_{\eta_i-\xi_j-i+j}(u-2\mu'_{\mu_1}+2r)    
              \,\,\,     i= \eta'_1-\mu'_1+\mu'_{\mu_1}  \cr}\cr
&  &(4.9{\rm b})\cr
{\cal H}^R_m(u) &= \sum_{l=0}^{m} (-1)^{l}
               T^{(r)}_1 (u-2l) T_{m-l}(u-m-r-l+{1 \over 2}).
&(4.9{\rm c})}
$$
\par
From (4.8a), (4.9a,c) and Theorem 3.1, one sees that
${\cal H}^R_m(u)$ is equal to the R-hatched ``dual hook''
$S^R_{(m^{r-1}) \subset ((m+1)^r)}(u)$.
From Theorems 2.1, 4.1 and 4.2, we have
\proclaim Corollary.
$S^L_{\lambda \subset \mu}(u)$ and
$S^R_{\lambda \subset \mu}(u)$ are pole-free provided that the BAE
(2.1) holds.
\par
Following a similar argument to the previous section, we propose the 
Drinfeld polynomials corresponding to the auxiliary
spaces $W^L_{\lambda \subset \mu}(u)$ and
$W^R_{\lambda \subset \mu}(u)$ of 
$S^L_{\lambda \subset \mu}(u)$ and 
$S^R_{\lambda \subset \mu}(u)$, respectively.
$$
P^{W^L_{\lambda \subset \mu}(u)}_a(\zeta)
 = P^{W_{\lambda \subset \mu}
(u+\mu^\prime_1+\mu_1-r-{3 \over 2})}_a(\zeta)
\quad 1 \le a \le r-1,\eqno(4.10{\rm a})
$$
$$\eqalign{
P^{W^L_{\lambda \subset \mu}(u)}_r(\zeta)
&= {1 \over \zeta - u + 1} P^{W_{\lambda \subset \mu}
(u+\mu^\prime_1+\mu_1-r-{3 \over 2})}_r(\zeta)\cr
&= \prod_{ \scriptstyle 2 \le i \le \mu_1  \atop
                \scriptstyle \lambda'_i+r \le \mu'_i \le \lambda'_{i-1}+r }
            (\zeta-u+3+2(\mu'_i-i-\mu^\prime_1))\cr
& \times \prod_{ \scriptstyle 1 \le i \le \mu_1 \atop
                \scriptstyle  \mu^\prime_{i+1}\le \lambda'_i+r \le \mu'_i }
           (\zeta-u+2+2(\lambda'_i-i-\mu^\prime_1+r)),\cr}
\eqno(4.10{\rm b})
$$
$$
P^{W^R_{\lambda \subset \mu}(u)}_a(\zeta)
 = P^{W_{\lambda \subset \mu}
(u-\mu^\prime_1-\mu_1+r+{3 \over 2})}_a(\zeta)
\quad 1 \le a \le r-1,\eqno(4.11{\rm a})
$$
$$\eqalign{
P^{W^R_{\lambda \subset \mu}(u)}_r(\zeta)
&= {1 \over \zeta - u - 1} P^{W_{\lambda \subset \mu}
(u-\mu^\prime_1-\mu_1+r+{3 \over 2})}_r(\zeta)\cr
&= \prod_{ \scriptstyle 1 \le i \le \mu_1  \atop
                \scriptstyle \lambda'_i+r \le \mu'_i \le \lambda'_{i-1}+r }
            (\zeta-u+2(\mu'_i-i+\mu_1-r))\cr
& \times \prod_{ \scriptstyle 1 \le i \le \mu_1-1 \atop
                \scriptstyle  \mu^\prime_{i+1}\le \lambda'_i+r \le \mu'_i }
           (\zeta-u-1+2(\lambda'_i-i+\mu_1)),\cr}
\eqno(4.11{\rm b})
$$
where we assume (3.12).
\par
In $Atab_L(\lambda \subset \mu)$
and $Atab_R(\lambda \subset \mu)$, 
we have considered the hatched part ($Spin$ (4.1))
only in the bottom left or top right position.
A natural question may be whether it is possible to define 
a tableau sum that becomes pole-free and 
contains $Spin$ simultaneously in various places 
in a skew-Young diagram $\lambda \subset \mu$.
It is indeed possible 
to include $Spin$ both at the bottom left and the top right.
However, we have found only few examples beyond that so far.
%


\beginsection 5. Solution to the $T$-system

\noindent
The functions 
$T_{\lambda \subset \mu}(u)$ (3.2), 
$S^L_{\lambda \subset \mu}(u)$ (4.3) and 
$S^R_{\lambda \subset \mu}(u)$ (4.5) provide
the solution to the $T$-system for $B_r$, one of the 
functional relations proposed in [KNS] for any $X_r$.
(See [KS2] for the $T$-system of twisted quantum affine algebras.)
For $m \in {\bf Z}_{\ge 0}$, put
$$\eqalignno{
T^{(a)}_m(u) &= T_{(m^a)}(u) \quad 1 \le a \le r-1,
&(5.1{\rm a})\cr
T^{(r)}_{2m}(u) &= T_{(m^r)}(u),&(5.1{\rm b})\cr
T^{(r)}_{2m+1}(u) &= S^L_{((m+1)^r)}(u-m) =
S^R_{((m+1)^r)}(u+m).&(5.1{\rm c})\cr}
$$
The latter equality in (5.1c) can be shown easily 
by using 
(2.14), (4.7a) and (4.9a).
The definition (5.1) includes (2.10).
Moreover, from (3.11) and (4.10,11),
the Drinfeld polynomial corresponding to
$T^{(a)}_m(u)$ is given by
$P_b(\zeta)=
\bigl(\prod_{i=1}^m(\zeta-u+{m+1-2i\over t_a})\bigr)^{\delta_{b a}}$
for $1 \le b \le r$, 
in agreement with (2.3) of [KS1].
Thus $T^{(a)}_m(u)$ here is the DVF for the transfer matrix 
$T^{(a)}_m(u)$ considered in [KS1].
\proclaim Theorem 5.1.
$T^{(a)}_m(u)$ defined above satisfies the following functional relations.
$$\eqalign{
T^{(a)}_m(u-1)T^{(a)}_m(u+1) &=
T^{(a)}_{m+1}(u)T^{(a)}_{m-1}(u) +
T^{(a-1)}_m(u)T^{(a+1)}_m(u)\cr
&\qquad\qquad\qquad\qquad\hbox{ for }\, 1 \le a \le r-2,\cr
T^{(r-1)}_m(u-1)T^{(r-1)}_m(u+1) &=
T^{(r-1)}_{m+1}(u)T^{(r-1)}_{m-1}(u) + 
T^{(r-2)}_m(u)T^{(r)}_{2m}(u),\cr
T^{(r)}_{2m}(u-{1\over 2})T^{(r)}_{2m}(u+{1\over 2}) &=
T^{(r)}_{2m+1}(u)T^{(r)}_{2m-1}(u) \cr
&+ T^{(r-1)}_m(u-{1\over 2})T^{(r-1)}_m(u+{1\over 2}),\cr
T^{(r)}_{2m+1}(u-{1\over 2})T^{(r)}_{2m+1}(u+{1\over 2}) &=
T^{(r)}_{2m+2}(u)T^{(r)}_{2m}(u) +
T^{(r-1)}_m(u)T^{(r-1)}_{m+1}(u).\cr}
\eqno(5.2)
$$
\par 
{\it Outline of the proof.} 
We use the determinantal expressions (3.5a) and (4.6a).
Then the first two equations in (5.2) reduce to the Jacobi
identity.
(cf. [KNS] eqs.(2.20)-(2.22).)
To prove the third equation, substitute (4.6a) into
$T^{(r)}_{2m+1}(u)T^{(r)}_{2m-1}(u)$.
Expanding the determinants with respect to the first column,
we have
$$\eqalign{
&T^{(r)}_{2m+1}(u)T^{(r)}_{2m-1}(u) =
\sum_{i=0}^{m-1}\sum_{j=0}^m (-1)^{i+j}
R^{(m)}_jR^{(m-1)}_i\cr
&\times \bigl(T^{r+j-i-1}(u-m+i+j+{1\over 2})+
T^{r+i-j}(u-m+i+j+{1\over 2})\bigr).\cr}
$$
Here, $R^{(m)}_j$ denotes the cofactor of 
$T^{(r)}_1(u-m+2j)$ in $T^{(r)}_{2m+1}(u)$
and we have used (2.14).
On taking the $j$-sum, the 
$T^{r+j-i-1}(u-m+i+j+{1\over 2})$ term vanishes.
After taking the $i$-sum, the 
$T^{r+i-j}(u-m+i+j+{1\over 2})$ term is non-zero
only for $j=0$ or $j=m$.
Noting that 
$R^{(m)}_0 = T^{(r)}_{2m}(u+{1\over 2})$
and $R^{(m)}_m = T^{(r-1)}_m(u-{1\over 2})$,
one has the third equation.
The last equation in (5.2) can be verified quite similarly.
\par
The functional relation (5.2) is the unrestricted $T$-system for $B_r$, 
(3.20) in [KNS] (in a different normalization).
There was a factor $g^{(a)}_m(u)$ 
in each equation as
$T^{(a)}_m(u+{1 \over t_a})
T^{(a)}_m(u-{1 \over t_a}) = 
T^{(a)}_{m+1}(u)T^{(a)}_{m-1}(u) + g^{(a)}_m(u)(\cdots)$.
The $g^{(a)}_m(u)$ 
is 1 here because
we are considering the case where vacuum part = 1.
The choice (5.1a) has been conjectured 
in (4.20) of [KS1] including the vacuum parts.
The case $r = 2$ had been proved earlier [K].
\par
It may be interesting to regard $u$ and $m$ as discrete space-time
variables and consider (5.2) as a 
discretized Toda equation.
Actually, a ``continuum limit'' of (5.2) (with 
$g^{(a)}_m(u)$) under an 
appropriate rescaling of $u, m$ and $g^{(a)}_m(u)$
leads to
$$
(\partial_u^2 - \partial_m^2)\hbox{ log } \phi_a(u,m)
= \hbox{const } \prod_{b=1}^r
\phi_b(u,m)^{-A_{a b}},$$
where $\phi_a(u,m)$ is a scaled $T^{(a)}_m(u)$
and $A_{a b} = {2(\alpha_a \vert \alpha_b)\over
(\alpha_a \vert \alpha_a)}$ is the Cartan matrix.
The constant above can be made arbitrary by choosing
the $g^{(a)}_m(u)$ suitably.
We remark that 
the $T$-system proposed in [KNS] has this aspect
for all the classical simple Lie algebra $X_r$.


\beginsection 6. On vacuum parts and BAE in terms of Drinfeld polynomial

\noindent
So far we have treated the case where the quantum space is
formally trivial.
This corresponds to choosing the LHS of the BAE (2.1) to be just -1
and the vacuum parts in the DVFs
$T_{\lambda \subset \mu}(u)$,
$S^L_{\lambda \subset \mu}(u)$ and 
$S^R_{\lambda \subset \mu}(u)$ to be 1.
To recover the vacuum parts for the non-trivial quantum space
$$
\otimes_{i=1}^N W^{(i)},\eqno(6.1)
$$
one needs to know the corresponding BAE.
Assuming that each $W^{(i)}$ in (6.1) is a finite dimensional 
irreducible $Y(B_r)$ module characterized by the Drinfeld
polynomial $P^{(i)}_a(\zeta) \, (1 \le a \le r)$,
we conjecture the BAE:
$$\eqalign{
-{P_a(v^{(a)}_k+{1\over t_a}) \over
P_a(v^{(a)}_k-{1\over t_a}) } &=
 \prod_{b = 1}^r 
{Q_b(v^{(a)}_k + (\alpha_a \vert \alpha_b)) \over
 Q_b(v^{(a)}_k - (\alpha_a \vert \alpha_b)) }
\quad 1 \le a \le r, \, 1 \le k \le N_a,\cr
P_a(\zeta) &= \prod_{i=1}^N P^{(i)}_a(\zeta).\cr}
\eqno(6.2)
$$
Here we understand that $q \rightarrow 1$ in (2.2)
for $Y(B_r)$.
(On the other hand, for generic $q$, we suppose 
that (6.2) is the correct BAE for $U_q(B^{(1)}_r)$
if $P^{(i)}_a(\zeta)$ is replaced by a natural $q$-analogue.)
The equation (6.2) has been formulated purely from the 
representation theoretical data, the root system 
and the Drinfeld polynomial.
Thus we suppose that it is the BAE for any $Y(X_r)$
(or $U_q(X^{(1)}_r)$ in the trigonometric case).
This is actually true for all the known examples in 
which alternative derivations of the BAE are known such as
the algebraic Bethe ansatz.
It is also agreed in [ST].
Once (6.2) is admitted, 
the vacuum parts are determined uniquely up to an overall scalar
by requiring that the pole-freeness
is ensured by (6.2).
This is a straightforward task and here we shall only 
indicate the initial step concerning Theorems 2.1 and 2.3.
\par
Redefine $z(a;u)$ (2.4) and 
$sp(\xi_1,\ldots,\xi_r;u)$ (2.5) by multiplying 
the vacuum parts $vac (\cdots)$
(cf. (2.9a) in [KS1]):
$$\eqalign{
vac\, z(a;u) &= \prod_{j=1}^{a-1}P_j(u+j-1)
\prod_{j=a}^{r-1}P_j(u+j+1) 
P_r(u+r+{1\over 2})P_r(u+r-{1\over 2})\cr
&\times \prod_{j=1}^{r-1}P_j(u+2r-j) \Phi(u)\quad 1 \le a \le r,\cr
vac\, z(0;u) &= \prod_{j=1}^{r-1}P_j(u+j-1)
P_r(u+r-{1\over 2})^2
\prod_{j=1}^{r-1}P_j(u+2r-j) \Phi(u),\cr
vac\, z({\bar a};u) &= \prod_{j=1}^{r-1}P_j(u+j-1)
P_r(u+r-{1\over 2})P_r(u+r-{3\over 2})\cr
&\times \prod_{j=1}^{a-1}P_j(u+2r-j)
\prod_{j=a}^{r-1}P_j(u+2r-j-2) \Phi(u)\quad 1 \le a \le r,\cr
}\eqno(6.3{\rm a})$$
$$
\Phi(u) = \prod_{b=1}^r \prod_{j=1}^{b-1}
P_b(u+b-2j-{1\over t_b})
P_b(u+2r-b+2j-1+{1\over t_b}).\eqno(6.3{\rm b})
$$
$$\eqalignno{
vac\, sp(\xi_1,\ldots, \xi_r; u) &= 
\psi^{(1)}_{n_1}(u) \cdots \psi^{(r)}_{n_r}(u), &(6.4{\rm a})\cr
n_b &= \sharp \{ j \mid \xi_j = -, 1 \le j \le b \},
&(6.4{\rm b})\cr
\psi^{(b)}_n(u) &= \prod_{j=0}^{n-1}
P_b(u+r-b+2j+{1\over 2}-{1\over t_b})\cr
&\times 
\prod_{j=n}^{b-1}
P_b(u+r-b+2j+{1\over 2}+{1\over t_b}).&(6.4{\rm c})\cr}
$$
In terms of $z(a;u)$ involving the above vacuum parts,
redefine $T^a(u)$ by (2.7a)
assuming $X P_b(u) = P_b(u+2) X\, (1 \le b \le r)$ and modifying the 
RHS into
$$\eqalign{
&\sum_{a=0}^\infty
F_a(u+a-1) T^a(u+a-1) X^a,\cr
&F_a(u) = \prod_{b=1}^r \prod_{j=1}^{a-1}
\psi_0^{(b)}(u+r-a-{1\over 2}+2j)
\psi_b^{(b)}(u-r+a+{1\over 2}-2j).\cr}
$$
It is easily seen that this $T^a(u)$ is of positive order $2b$ 
with respect to the $P_b$ function (6.2).
One can check that Theorem 2.1 is still valid 
(for $T^{(r)}_1(u)$ and $T^a(u)$) for the BAE (6.2).
Relations (2.12a) and (2.13a) also hold if 
the right hand sides are divided by
$F_a(u)$ and $F_{2r-1-a}(u)$, respectively.
Thus Theorem 2.3 remains valid without any changes.
Along these lines, one can proceed further 
to include the vacuum parts for general
$T_{\lambda \subset \mu}(u)$, 
$S^L_{\lambda \subset \mu}(u)$ and 
$S^R_{\lambda \subset \mu}(u)$
so that they become pole-free under the BAE (6.2).

%
%
\beginsection Acknowledgments

The authors thank 
E. Date, R. Hirota, J. Satsuma,
E.K. Sklyanin, V.O. Tarasov and I. Terada for discussions.
\beginsection References

\item{[BR]}{V.V. Bazhanov,V.V., N.Yu. Reshetikhin,
J.Phys.A:Math.Gen. {\bf 23} (1990) 1477}
\item{[C]}{I. Cherednik, in Proc. of the XVII International 
Conference on Differential Geometric Methods in Theoretical Physics,
Chester, ed. A.I. Solomon, World Scientific, Singapore, 1989}
\item{[D]}{V.G. Drinfel'd,
Sov.Math.Dokl.{\bf 36} (1988) 212}
\item{[K]}{A. Kuniba, J. Phys.A: Math.Gen.{\bf 27} (1994) L113}
\item{[KN]}{M. Kashiwara and T. Nakashima, ``Crystal graphs for 
representations of the}
\item{}{$q$-analogue of classical Lie algebras.''
RIMS preprint {\bf 767} (1991)}
\item{[KNS]}{A. Kuniba, T. Nakanishi and J. Suzuki,
Int.J.Mod.Phys. {\bf A9} (1994) 5215}
\item{[KS1]}{A. Kuniba and J. Suzuki, 
``Analytic Bethe ansatz for fundamental representations of 
Yangians'', hep-th.9406180, Commun. Math. Phys. in press}
\item{[KS2]}{A. Kuniba and J. Suzuki, J. Phys.A: Math.Gen.
{\bf 28} (1995) 711}
\item{[KT]}{K. Koike and I. Terada, J. Alg. {\bf 107} (1987) 466}
\item{[M]}{I. G. Macdonald, {\it Symmetric functions and Hall polynomials},
2nd ed., Oxford University Press, 1995}
\item{[R]}{N.Yu. Reshetikhin, Sov.Phys.JETP {\bf 57} (1983) 691,
Theor.Math.Phys.{\bf 63} (1985) 555, 
Lett.Math.Phys.{\bf 14} (1987) 235}
\item{[RW]}{N.Yu. Reshetikhin and P.B. Wiegmann,
Phys.Lett.B{\bf 189} (1987) 125}
\item{[ST]}{E.K. Sklyanin and V.O. Tarasov, private communication}

\vfill\eject

\beginsection Figure Captions.

\par\noindent 
Figure 3.1:  An example of a skew-Young diagram 
$\lambda \subset \mu$. Here $\mu=(5,4^2,1),
\lambda=(2,1), \mu'=(4,3^3,1)$ and $\lambda'=(2,1)$, respectively.
The lengths of the main diagonal are given by $d_{\mu}=3$ and 
$d_{\lambda}=1$.
\par\noindent
Figure 3.2:  The way to assign the letters to each box 
is explained in the text.
This is an example for $r=3, \mu'=(9,7,2), \lambda'=(3,1)$. Notice that
zeros are arranged lest they are adjacent horizontally. 
\par\noindent
Figure 4.1: An example of an L-hatched skew-Young diagram
$r=4, \mu=(4^3,3,2,1^2), \lambda=(3,1)$.
\par\noindent
Figure 4.2: The bottom left part of an
L-hatched skew-Young tableau and the assignment of 
the letters $\{i_l\}$ and $\{j_l\}$ in (4.2).
\par\noindent
Figure 4.3: An example of an R-hatched skew-Young diagram
$r=4, \mu=(4^5,3,1), \lambda=(3^2,2,1)$.
\par\noindent
Figure 4.4: The top right part of an
R-hatched skew-Young tableau and the assignment of the letters 
$\{i_l\}$ and $\{j_l\}$ in (4.4).
\par\noindent
Figure 4.5: An R-hatched skew-Young diagram for $r=3$ with $\mu=(5^4,3,2,1),
\lambda=(4,3,1^2)$. Broken lines are guides to eyes for defining
$\eta=(4^3,3,2,1), \xi=(3, 1^2)$.

\bye